\documentclass[preprintnumbers,nofootinbib,noshowpacs,eqsecnum,prd]{revtex4}

\usepackage{graphicx,epsfig}
\usepackage{amsmath,amssymb,bbold}
\usepackage{url}
\usepackage{color}
\usepackage{axodraw}

\graphicspath{ {Plots/} }

\makeatletter       

\renewcommand{\p@subsection}{}

\makeatother

\DeclareMathOperator{\tr}{tr}

\newcommand{\qw}{\ensuremath{\tilde{q}}}

\newcommand{\gw}{\ensuremath{\tilde{g}}}

\begin{document}
\def\lsim{\:\raisebox{-0.5ex}{$\stackrel{\textstyle<}{\sim}$}\:}
\def\gsim{\:\raisebox{-0.5ex}{$\stackrel{\textstyle>}{\sim}$}\:}

\title{ \hfill{\footnotesize{CERN-PH-TH/2008-237}}      \\
        \hfill{\footnotesize{DESY 08-188}}      \\
        \hfill{\footnotesize{LPT-ORSAY 08-103}} \\
        \hfill{\footnotesize{PITHA 08/31}}     \\[3mm]
        COLOR-OCTET SCALARS OF N=2 SUPERSYMMETRY AT THE LHC \\[1mm]
}

\author{S.~Y.~Choi$^1$, M.~Drees$^{2}$, J.~Kalinowski$^3$,
        J.~M.~Kim$^2$, E.~Popenda$^4$ and P.~M.~Zerwas$^{4,5}$ \\[-3mm]
        \mbox{ } }
\address{ $^1$ Department of Physics and RIPC, Chonbuk National University,
               Jeonju 561-756, Korea \\
          $^2$ Physikalisches Inst.~der Univ.~Bonn, D-53115 Bonn, Germany and
             \\ Bethe Center for Theoretical Physics, Univ. Bonn, D-53115 Bonn,
               Germany \\
          $^3$ Physics Department, University of Warsaw, 00681 Warsaw, Poland and \\
               Theory Division, CERN, CH-1211 Geneva 23, Switzerland\\
          $^4$ Inst.~Theor.~Physik E, RWTH Aachen U, D-52074 Aachen, Germany \\
          $^5$ Deutsches Elektronen-Synchrotron DESY, D-22603 Hamburg, Germany and \\
               Laboratoire de Physique Th\'{e}orique, U. Paris-Sud, F-91405 Orsay,
               France }

\date{\today}

\begin{abstract}
  {\it The color gauge hyper-multiplet in N=2 supersymmetry consists of the
    usual N=1 gauge vector/gaugino super-multiplet, joined with a novel
    gaugino/scalar super-multiplet. Large cross sections are predicted for the
    production of pairs of the color-octet scalars $\sigma$ [sgluons] at the
    LHC: $gg, q\bar{q} \to \sigma\sigma^{\ast}$. Single $\sigma$ production is
    possible at one-loop level, but the $g g\to \sigma$ amplitude vanishes in
    the limit of degenerate $L$ and $R$ squarks. When kinematically allowed,
    $\sigma$ decays predominantly into two gluinos, whose cascade decays give
    rise to a burst of eight or more jets together with four LSP's as
    signature for $\sigma$ pair events at the LHC.  $\sigma$ can also decay
    into a squark-antisquark pair at tree level. At one-loop level $\sigma$
    decays into gluons or a $t \bar t$ pair are predicted, generating exciting
    resonance signatures in the final states. The corresponding partial widths
    are very roughly comparable to that for three body final states mediated
    by one virtual squark at tree level.}
\end{abstract}

\maketitle

\section{Introduction}
\label{sec:introduction}

\noindent
The pairwise production of supersymmetric squarks and gluinos at the LHC leads
to final states that contain two to four hard jets [plus somewhat softer jets
from QCD radiation and/or decays of heavier neutralinos and charginos] and
missing transverse momentum generated by two LSP's. These signatures are
typical for N=1 supersymmetry \cite{Wess,Nilles,Drees} as specified in the
Minimal Supersymmetric Standard Model [MSSM]. However, in alternative
realizations of supersymmetry the final-state topology could be rather
different. In order to exemplify this point, we have adopted an N=1/N=2 hybrid
model, cf.  Ref.~\cite{Fayet,CDFZ,Benakli}, in which supersymmetry
characteristics are quite different from the MSSM. Assuming the N=2 mirror
(s)fermions to be very heavy in order to avoid chirality problems, the hybrid
model expands to N=2 only in the gaugino sector. The QCD sector is built up by
the usual N=1 gluon/gluino super-multiplet, joined with an additional
gluino/scalar super-multiplet. [Similarly, the electroweak sector is
supplemented by additional SU(2)$_L$ and U(1)$_Y$ super-multiplets; this
sector will not be discussed here.] For the sake of simplicity we will
disregard in the analysis mass splittings of the scalar fields and we assume
equal masses for the usual and the novel gluinos which, as a result, can be
combined to a common Dirac field, see Refs.~\cite{Benakli,CDFZ,no}.  Since the
experimental consequences of variations involving a larger set of parameters
are rather obvious, they will not be discussed in this letter.

The novel scalar color-octet fields $\sigma$ [which may be called scalar
gluons\footnote{Not to be confused with the scalar gluons that were discussed
  as carriers of the strong force in alternatives to QCD constructed in the
  1970's.}, or contracted to sgluons \cite{TT}] can be produced
  in pairs:
\begin{eqnarray}
&&   gg,\; q\bar{q} \to \sigma \sigma^{\ast}   \,.
\label{eq:prod}
\end{eqnarray}
The color-octet sgluons are $R$-parity even, and thus can also be produced singly in
gluon-gluon or quark-antiquark collisions, albeit through loop processes only:
\begin{eqnarray}
&&   gg,\; q \bar q \to \sigma  \,.
\label{eq:prod1}
\end{eqnarray}
However, as we will show, the corresponding matrix elements vanish in the
limit of degenerate $L$ and $R$ squarks. Moreover, single sgluon production in
quark-antiquark collisions proceeds through a chirality-flip process that is
suppressed, strongly in practice, by the quark mass.

At tree level, $\sigma$ can decay either into (real or virtual) gluino or
squark pairs,
\begin{eqnarray} \label{sdec-tree}
&&   \sigma \to \tilde{g} \tilde{g} \to qq \tilde{q} \tilde{q}
            \to qqqq + \tilde{\chi} \tilde{\chi}\,,   \nonumber\\
&&   \sigma \to \tilde{q} \tilde{q} \to qq +  \tilde{\chi} \tilde{\chi} \,,
\label{eq:sdec-mod}
\end{eqnarray}
where $\tilde \chi$ denotes electroweak neutralinos or charginos. At
one-loop level, $\sigma$ can also decay into top-quark or gluon pairs:
\begin{eqnarray} \label{sdec-loop}
&& \sigma \to t \bar t \to b \bar b W^+ W^-\,,\nonumber\\
&& \sigma \to gg\,.
\end{eqnarray}
Apart from the last mode, these lead to spectacular signatures for $\sigma$
pair production at the LHC, e.g.
\begin{eqnarray} \label{signatures}
   pp &\to& 8\, {\rm{jets}} + 4\, {\rm{LSP's}}\,,      \nonumber\\
   pp &\to& tt{\bar{t}}{\bar{t}}                                 \,.
\end{eqnarray}
In the first case a burst of eight almost isotropically distributed hard jets
is generated in $\sigma$-pair production, even not counting QCD stray jets nor
possible $\tilde \chi$ decay products, and a large amount of missing energy.
Alternatively, four top (anti)quarks are predicted by the second mechanism.
These signatures are very different from the usual MSSM topologies and raise
exciting new experimental questions. Likewise, single $\sigma$ production
followed by gluon-pair decays generates novel resonance signatures foreign to
N=1 supersymmetry.

Apart from Ref.~\cite{TT}, the possibility that there might exist SU(3)$_C$
octet scalars within reach of the LHC has recently been discussed in different
context in Refs.~\cite{others}. While the tree-level cross sections for the
pair production of these scalars at the LHC are the same in all these
scenarios [up to trivial multiplicity factors], the possibilities of single
production, as well as the decay modes and experimental signatures of the
scalars in both channels, are quite different in our case and Ref.~\cite{TT}
from those discussed earlier.

This note is divided into two parts. In the next section the theoretical basis
of the N=1/N=2 hybrid model will be recapitulated briefly, and the
loop-induced $\sigma gg$ and $\sigma q \bar q$ couplings will be discussed.
The third section is devoted to the phenomenology of $\sigma$-pair production
and cascade decays, followed by a short analysis of single $\sigma$ production
in gluon fusion.

\section{THEORETICAL BASIS: GAUGE HYPER-MULTIPLETS  AND SCALARS}
\label{sec:theoretical_base}

\noindent
As noted earlier, the N=2 QCD hyper-multiplet can be decomposed into the usual
N=1 octet gluon/gluino multiplet $\hat{g}=\{g_\mu, \tilde{g} \}$ plus an N=1
octet multiplet $\hat{g}'=\{\sigma,\tilde{g}'\}$ of extra gluinos and scalar
$\sigma$ fields. Schematically, the QCD hyper-multiplet is described by a
diamond plot,
%
\begin{center}
\begin{picture}(150,80)(0,15)
\Text(70,80)[c]{\color{black} $g_\mu$}
\Text(30,50)[c]{\color{black} $\tilde{g}$}
\Text(110,50)[c]{\color{black} $\tilde{g}'$}
\Text(70,20)[c]{\color{black} $\sigma$}
\SetWidth{0.5}
\Line(35,55)(62,76)
\Line(75,25)(102,46)
\Text(150,100)[c]{Spin}
\Text(150,80)[c]{\color{black} $1$}
\Text(150,50)[c]{\color{black} $1/2$}
\Text(150,20)[c]{\color{black} $0$}
\end{picture}
\end{center}
%
where the first, second and third row corresponds to spin 1, 1/2 and 0 states.
The N=1 superfields are represented by the two pairs connected by the thin
lines. The $\sigma$ field carries positive $R$-parity.

The only gauge invariant term in the N=1 superpotential containing the new
gluino/sgluon superfield $\hat{g}'$ is a mass term,
\begin{equation} \label{W}
W_{\hat g'}\, =\, \frac{1}{2}\, M_3'\, \hat g'^a \hat g'^a\,,
\end{equation}
where we have adopted the notation of Ref.$\,$\cite{CDFZ}. The only
supersymmetric interactions involving $\hat g'$ are thus QCD gauge
interactions plus gauge strength $\sigma \tilde g \tilde g' $ Yukawa-type
interactions \cite{CDFZ}. In a full N=2 theory, there would also be couplings
between $\hat g'$ and the N=2 partners of the usual matter superfields;
however, in our hybrid construction we assume the latter to be decoupled
from TeV scale physics.

The masses of the new scalars are determined by the superpotential (\ref{W})
plus soft breaking terms \cite{fnw}
\begin{equation} \label{L_soft}
{\cal L}_{\sigma,\,{\rm soft}} = -m_\sigma^2 \left| \sigma^2 \right| -
  \left(m^2_{\sigma\sigma}  \sigma \sigma + {\rm h.c.} \right) -
 g_s M_3^D \left[ \sigma^a \frac{\lambda^a_{ij}} {\sqrt{2}}
  \sum_q \left( \tilde q_{Li}^* \tilde q_{Lj} - \tilde q_{Ri}^* \tilde q_{Rj}
  \right) + {\rm h.c.} \right] \,,
\end{equation}
where $g_s$ is the strong coupling constant and $\lambda^a$ are the Gell-Mann
SU(3)$_C$ matrices. The parameter $M_3^D$ is the Dirac gluino mass connecting
$\tilde g'$ with the usual gluino $\tilde g$~\cite{CDFZ}. If the supersymmetry breaking is
spontaneous, the Dirac gluino mass also gives rise to a supersymmetry breaking
trilinear scalar interaction between $\sigma$ and the MSSM squarks, as shown
in Eq.$\,$(\ref{L_soft}); note that $L$ and $R$ squarks contribute with
opposite signs as demanded by the general form of the super-QCD $D$-terms
[differing from the first version of Ref.$\,$\cite{TT} with far reaching
phenomenological consequences].\footnote{If one allows oneself the freedom to
  break supersymmetry explicitly, but softly, the coefficients of the $\sigma
  \tilde q \tilde q$ interactions would be arbitrary, and could even be set to
  zero; this would, however, not be stable against radiative corrections.} As
noted above, we will set $m^2_{\sigma\sigma}= 0$ in this discussion, so that
the physical mass of the complex scalar octet is
\begin{equation} \label{mass}
M_\sigma = \sqrt{\left| M_3' \right|^2 + m^2_\sigma} \,.
\end{equation}
For given mean mass, a nonzero $m_{\sigma\sigma}^2$ generating a mass
splitting of the scalar fields would increase the total cross section for the
production of the new scalars.

In the simplest realization the two gluinos, $\tilde{g}$ and $\tilde{g}'$, are
not endowed with individual masses [i.e. $M_3' = 0$] but they are coupled by
the mass parameter $M_3^D$ in a purely off-diagonal mass matrix.\footnote{Note
  that this Dirac mass term must be nonzero, since otherwise the lightest
  member of the superfield $\hat g'$ would be stable. In contrast, scenarios
  where the diagonal Majorana entries of the gluino mass matrix vanish are
  perfectly acceptable.} In this configuration the two Majorana gluinos can be
combined to a 4-component Dirac gluino field $\tilde{g}_D$ as
\begin{eqnarray}
\tilde{g}_D = \tilde{g}_R + \tilde{g}'_L  \,,
\end{eqnarray}
with the mass eigenvalue given by $|M_3^D|$, cf. Ref.~\cite{CDFZ}. The
couplings of this Dirac field $\tilde{g}_D$ to the $\sigma$-field and to the
squark and quark fields are summarized in the interaction Lagrangians
\begin{eqnarray}
 {\cal L}_{\tilde{g}_D\tilde{g}_D\sigma}
   &=& -\sqrt{2} i\, g_s\, f^{abc}\, \overline{\tilde{g}^a_{DL}}\,
        \tilde{g}^b_{DR}\, \sigma^c
       +{\rm h.c.}\,, \\[1mm]
 {\cal L}_{\tilde{g}_D q \tilde{q}}
   &=& -\sqrt{2}\, g_s \sum_q
       \left( \overline{q_L} \frac{\lambda^a}{2}
              \tilde{g}^a_{DR}\,\, \tilde{q}_L
            + \overline{q_R} \frac{\lambda^{aT}}{2}
              \tilde{g}^{aC}_{DL}\,\, \tilde{q}_R\right)
              + {\rm h.c} \,.
\label{eq:lagrangian}
\end{eqnarray}
where $\tilde{g}^{CT}_D=-(\tilde{g}'_R + \tilde{g}_L)$ is the charge-conjugate
4-component Dirac gluino \cite{CDFZ}, $f^{abc}$ are the SU(3)$_C$ structure constants
and $\lambda^a$ are the Gell-Mann matrices. In addition, the sgluon fields couple to
gluons in tri-and quattro-linear vertices as prescribed by gauge theories for
scalar octet fields, i.e. proportional to the octet self-adjoint SU(3)$_C$
representation $F$. As a result, at tree level $\sigma$ pairs can be produced
in gluon collisions as well as in $q \bar q$ annihilation, but single production
of $\sigma$'s is not possible.

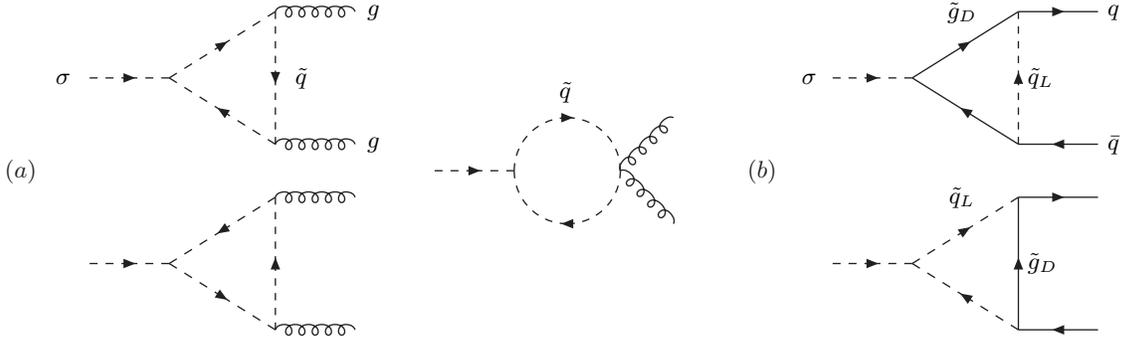
\begin{figure}[t]
\begin{center}
\begin{picture}(400,125)(0,0)
\Text(0,60)[r]{$(a)$}
\Text(10,95)[c]{\color{black} $\sigma$}
\Text(100,95)[c]{\color{black} $\tilde{q}$}
\DashArrowLine(20,95)(50,95){3}
\DashArrowLine(50,95)(90,120){3}
\DashArrowLine(90,70)(50,95){3}
\DashArrowLine(90,120)(90,70){3}
\Text(125,120)[l]{\color{black} $g$}
\Text(125,70)[l]{\color{black} $g$}
\Gluon(90,120)(120,120){2}{5}
\Gluon(90,70)(120,70){2}{5}
\DashArrowLine(20,25)(50,25){3}
\DashArrowLine(90,50)(50,25){3}
\DashArrowLine(50,25)(90,0){3}
\DashArrowLine(90,0)(90,50){3}
\Gluon(90,50)(120,50){2}{5}
\Gluon(90,0)(120,0){2}{5}
\Text(200,90)[c]{\color{black} $\tilde{q}$}
\DashArrowLine(150,60)(180,60){3}
\DashArrowArcn(200,60)(20,180,0){3}
\DashArrowArcn(200,60)(20,0,180){3}
\Gluon(220,60)(240,80){2}{4}
\Gluon(220,60)(240,40){2}{4}
\Text(280,60)[r]{$(b)$}
\Text(295,95)[r]{\color{black} $\sigma$}
\Text(350,120)[c]{\color{black} $\tilde{g}_D$}
\Text(375,95)[l]{\color{black} $\tilde{q}_L$}
\DashArrowLine(300,95)(330,95){3}
\ArrowLine(330,95)(370,120)
\ArrowLine(370,70)(330,95)
\DashArrowLine(370,70)(370,120){3}
\Text(405,120)[l]{\color{black} $q$}
\Text(405,70)[l]{\color{black} $\bar{q}$}
\ArrowLine(370,120)(400,120)
\ArrowLine(400,70)(370,70)
\Text(350,50)[c]{\color{black} $\tilde{q}_L$}
\Text(375,25)[l]{\color{black} $\tilde{g}_D$}
\DashArrowLine(300,25)(330,25){3}
\DashArrowLine(330,25)(370,50){3}
\DashArrowLine(370,0)(330,25){3}
\ArrowLine(370,0)(370,50)
\ArrowLine(370,50)(400,50)
\ArrowLine(400,0)(370,0)
\end{picture}
\end{center}
\caption{\it Diagrams for (a) the effective $\sigma gg$ vertex built up by
             squark loops; (b) the effective $\sigma q\bar{q}$ vertex with $L$
             squarks and gluinos -- the coupling to $R$ squarks being mediated
             by the charge-conjugate Dirac gluinos.}
\label{fig:loop_diagram}
\end{figure}

Even at the one-loop level, gluino loops do not contribute to the $\sigma gg $
coupling, due to the Bose symmetry of the gluons. The coupling is even in the
4-momenta under gluon exchange but it is odd, on the other hand, due to the
antisymmetric octet matrix elements $f^{abc}$ in color space. [Note that
SU(3)$_C$ singlet particles, like Higgs bosons, couple symmetrically to
gluons, by contrast.] Actually, the coupling of the octet sgluon to any number
of gluons is forbidden in the general softly broken N=2 pure gauge theory with
two Majorana gluinos [which may or may not be combined to a single Dirac
gluino] because the totally antisymmetric factor $f^{abc}$ forces the sgluon
to couple only to two {\em different} Majorana gluinos, while gluons always
couple to {\it diagonal} Majorana gluino pairs.

However, $\sigma$ can couple non-trivially to gluon pairs and quark-antiquark
pairs through triangle diagrams involving squark lines. Characteristic
examples are depicted in Fig.$\,$\ref{fig:loop_diagram}. In parallel to the
interaction Lagrangian it turns out that all $L$- and $R$-squark contributions
to the couplings come with opposite signs so that they cancel each other for
mass degenerate squarks. In addition, the quark-antiquark coupling is
suppressed by the quark mass as evident from general chirality rules.

\noindent
{\it{Comment.}} Before discussing the phenomenological implications,
let us note that the
presence of new fields in the N=1/N=2 hybrid model affects the
renormalization group (RG) running of gauge couplings above the weak scale; to
one-loop order,
\begin{eqnarray}
\frac{d \alpha^{-1}_i(Q^2)}{d \log(Q^2)} =  \frac{b_i}{2\pi} \,.
\end{eqnarray}
The coefficients $b_i$ for the non-Abelian group factors SU$(N_i)$
receive in the hybrid model contributions in addition to MSSM,
\begin{eqnarray}
b_i= b_i^{\rm MSSM} - \frac{2}{3} N_i - \frac{1}{3} N_i\,,
\end{eqnarray}
where the second term comes from the new Majorana fermions $\tilde{g}'$ and
the third from the complex scalars $\sigma$; the running of the U(1)$_Y$
coupling remains unaffected at one-loop order. As a result, gauge coupling
unification and the prediction of the weak mixing angle are lost; instead, the
couplings $g_i$ and $g_j$ meet at different points $M_{X,ij}$, all of which
lie above the Planck scale. Possible solutions to this problem would be to add
fields to the theory so that the new fields fall in complete GUT 
multiplets~\cite{Nelson:2002ca}, 
or to allow a different normalization for U(1)$_Y$ \cite{ael}, or to
contemplate different unification patterns \cite{fnw}, etc. Since in this
letter we are interested in the low-energy phenomenology of the color-octet
scalars, we will not delve into this subject any further.

\section{PHENOMENOLOGY OF COLOR-OCTET SCALARS AT THE LHC}
\label{sec:phenomenology}

\subsection{$\sigma$ Decays}

\noindent
At tree level the $\sigma$ particles can decay to a pair of Dirac gluinos
${\gw}_D$ or into a pair of squarks, with one or both of these sparticles
being potentially virtual when $M_{\sigma} < 2 M_{{\gw}_D}, 2 m_{\tilde q}$.
For on-shell decays and assuming pure Dirac gluinos the partial widths are
\begin{eqnarray} \label{Gamma_tree}
\Gamma [\sigma \to \gw_D {\bar{\gw}}_D ]
   &=& \frac{3 \alpha_s M_{\sigma}}{4}
        \beta_{\tilde{g}}\, (1+\beta^2_{\tilde{g}})\,, \nonumber \\
\Gamma[ \sigma \to \tilde{q}_a \tilde{q}_a^*] &=& \frac {\alpha_s}{4}
   \frac{|M_3^D|^2} {M_\sigma} \beta_{\tilde{q}_a}\,,
\end{eqnarray}
where $\beta_{\gw,\qw_a}$ are the velocities of $\gw,\qw_a$ ($a=L,R$).  In the
presence of non-trivial $\tilde q_L$-$\tilde q_R$ mixing the subscripts $L,R$
in the second Eq.(\ref{Gamma_tree}) have to be replaced by $1,2$ labeling the
mass eigenstates, and the contribution from this flavor is suppressed by a
factor $\cos^2 ( 2 \theta_{\qw})$; the mixing angle is defined via the
decomposition of the lighter mass eigenstate $\tilde q_1 = \cos \theta_{\qw}\,
\tilde q_L + \sin \theta_{\qw}\, \tilde q_R$. In addition, decays into $\tilde
q_1 \tilde q_2^*$ and $\tilde q_1^* \tilde q_2$ are possible, with the
coefficient $\sin^2(2 \theta_{\qw})$ and with the velocity $\beta_{\qw_a}$
replaced by the phase-space function $\lambda^{1/2}(1, m^2_{\tilde q_1} /
M^2_\sigma, m^2_{\tilde q_2} / M^2_\sigma)$. The gluinos subsequently decay to
quarks and squarks, again either real or virtual, and the squarks to quarks
and charginos/neutralinos tumbling eventually down to the LSP.

On the other hand, the trilinear interaction in Eq.$\,$(\ref{L_soft}) gives
rise to an effective $\sigma g g$ coupling via squark loops,
Fig.~\ref{fig:loop_diagram}(a), leading to the partial decay width
\begin{equation} \label{Gamma_gg}
\Gamma(\sigma \rightarrow g g)
  = \frac {5 \alpha_s^3} {384 \pi^2} \frac{|M_3^D|^2} {M_\sigma}
      \left| \sum_q \left[ \tau_{\tilde{q}_L} f(\tau_{\tilde{q}_L})
                         - \tau_{\tilde{q}_R} f(\tau_{\tilde{q}_R}) \right]
     \right|^2\,,
\end{equation}
with $\tau_{\tilde{q}_{L,R}} = 4 m^2_{\tilde q_{L,R}} / M^2_\sigma$ and
\cite{MS}
\begin{equation} \label{ftau}
f(\tau) = \left\{ \begin{array}{cl}
\left[ \sin^{-1} \left( \frac {1} {\sqrt{\tau}} \right) \right]^2  & {\rm
  for} \ \tau \geq 1\,, \\[3mm]
- \frac{1}{4} \left[ \ln \frac { 1 + \sqrt{1 - \tau} } { 1 - \sqrt{1 - \tau} }
  - i \pi \right]^2  \ & {\rm for}\ \tau < 1 \,.
\end{array} \right.
\end{equation}
In the presence of nontrivial $\tilde q_L$-$\tilde q_R$ mixing, the
subscripts $L,R$ in Eq.(\ref{Gamma_gg}) again have to be replaced by $1,2$
labeling the mass eigenstates, and the contribution from this flavor is
suppressed by a factor $\cos ( 2 \theta_{\qw})$ multiplying the term in square
parentheses. Note that the $\sigma g g$ coupling vanishes in the limit of
degenerate $L$ and $R$ squarks.

Furthermore, the $\sigma$ field couples to quark-antiquark pairs -- in
principle. By standard helicity arguments, this chirality-flip coupling is
suppressed however by the quark mass. For pure Dirac gluinos, the triangle diagrams,
Fig.~\ref{fig:loop_diagram}(b), either with two internal gluino lines and one
squark line or with two internal squark lines and one gluino line
again vanish for degenerate $L$ and $R$ squarks. The resulting partial width can
be written as
\begin{equation} \label{Gamma_qq}
\Gamma(\sigma \rightarrow q \bar q)
   = \frac{9 \alpha_s^3} {128 \pi^2} \frac{|M_3^D|^2 m_q^2}{M_\sigma}\,
     \beta_q
     \left[ \left(M^2_\sigma-4 m_q^2\right) |{\cal I}_S|^2
           +M^2_\sigma \, |{\cal I}_P|^2 \right]\,.
\end{equation}
The loop integrals for the effective scalar ($S$) and pseudoscalar ($P$)
couplings are given by
\begin{eqnarray} \label{I_loop}
{\cal I}_S
  &=& \int_0^1 dx \int_0^{1-x} dy \left\{ (1 - x- y)
       \left( \frac {1} {C_L} - \frac {1} {C_R} \right)
      +\frac{1}{9} (x+y)
       \left( \frac {1} {D_L} - \frac {1} {D_R} \right) \right\}
       \,, \nonumber \\
{\cal I}_P &=& \int_0^1 dx \int_0^{1-x} dy
  \left( \frac {1} {C_L} - \frac {1} {C_R} \right) \,,
\end{eqnarray}
where we have defined ($a= L,R$)
\begin{eqnarray} \label{C_loop}
C_a &=& (x+y) |M_3^D|^2 + (1-x-y) m^2_{\tilde q_a} - x y M^2_\sigma
 - (x+y) (1-x-y) m_q^2\,, \nonumber \\
D_a &=& (1-x-y) |M_3^D|^2 + (x+y) m^2_{\tilde q_a}
- x y M^2_\sigma - (x+y) (1-x-y) m_q^2\,.
\end{eqnarray}
${\cal I}_{S,P}$ can also be expressed in terms of standard Passarino-Veltman
functions \cite{pv}, e.g. ${\cal I}_P = C_{0L} - C_{0R}$, with
$C_{0L,R} \equiv C_0(|M_3^D|, m_{{\tilde q}_{L,R}}, |M_3^D|;
m_q^2, m_q^2, M^2_\sigma)$. In the presence of nontrivial $\tilde
q_L$-$\tilde q_R$ mixing, the subscripts $L,R$ in Eq.(\ref{I_loop}) have to
be replaced by $1,2$ labeling the squark mass eigenstates, and the
contribution from this flavor to the double integrals is suppressed by a
factor $\cos ( 2 \theta_q)$. Note that ${\cal I}_S = {\cal I}_P = 0$ if
$m_{\tilde q_L} = m_{\tilde q_R}$. In the presence of $\tilde{q}_L$-$\tilde{q}_R$
mixing this cancellation is no longer exact for two non-degenerate Majorana
gluinos.

\begin{figure}[bh!]
\vskip -0.3cm
\rotatebox{270}{\epsfig{figure=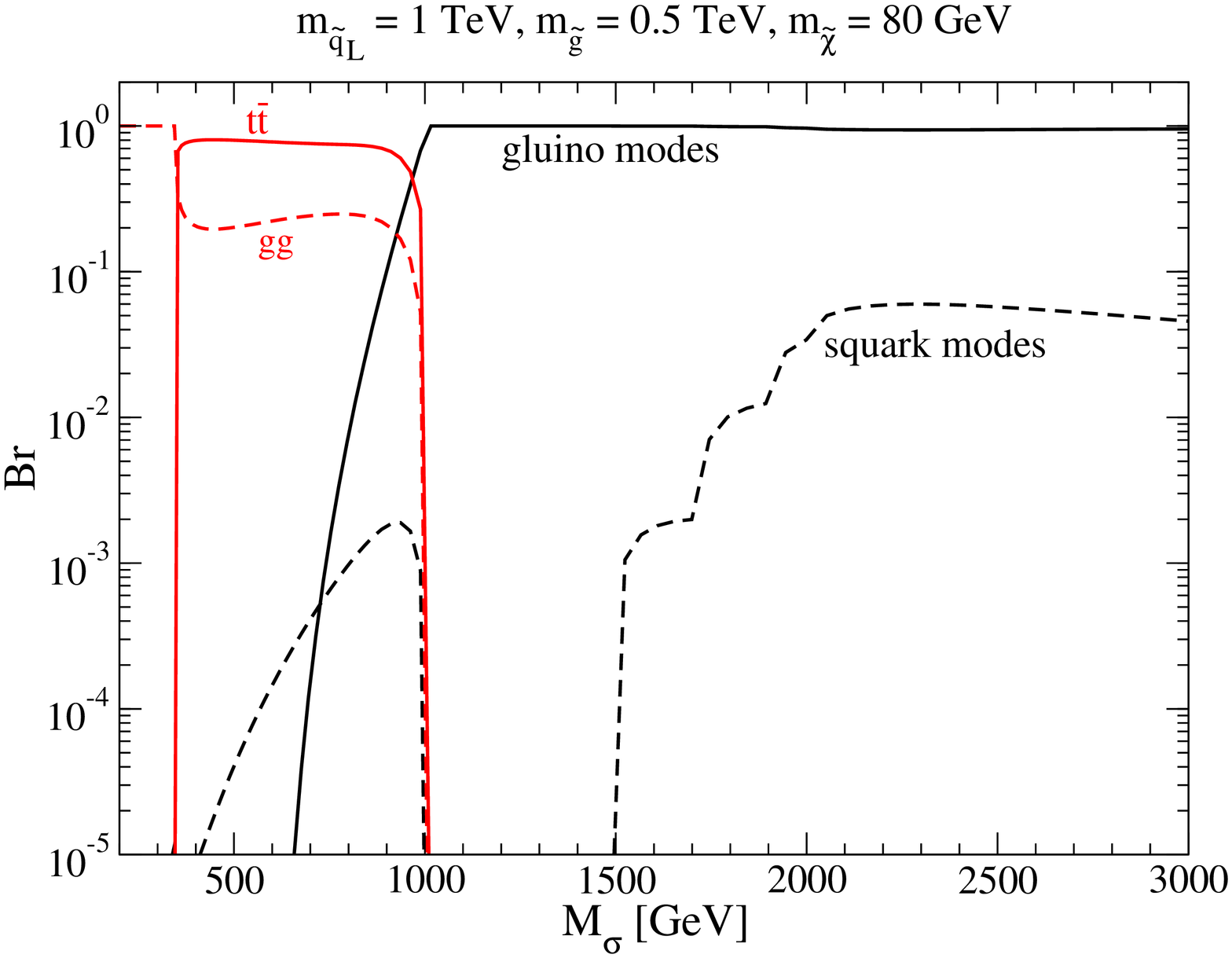, width=7.5cm,height=8.cm}}
\hspace{-0.5cm}
\rotatebox{270}{\epsfig{figure=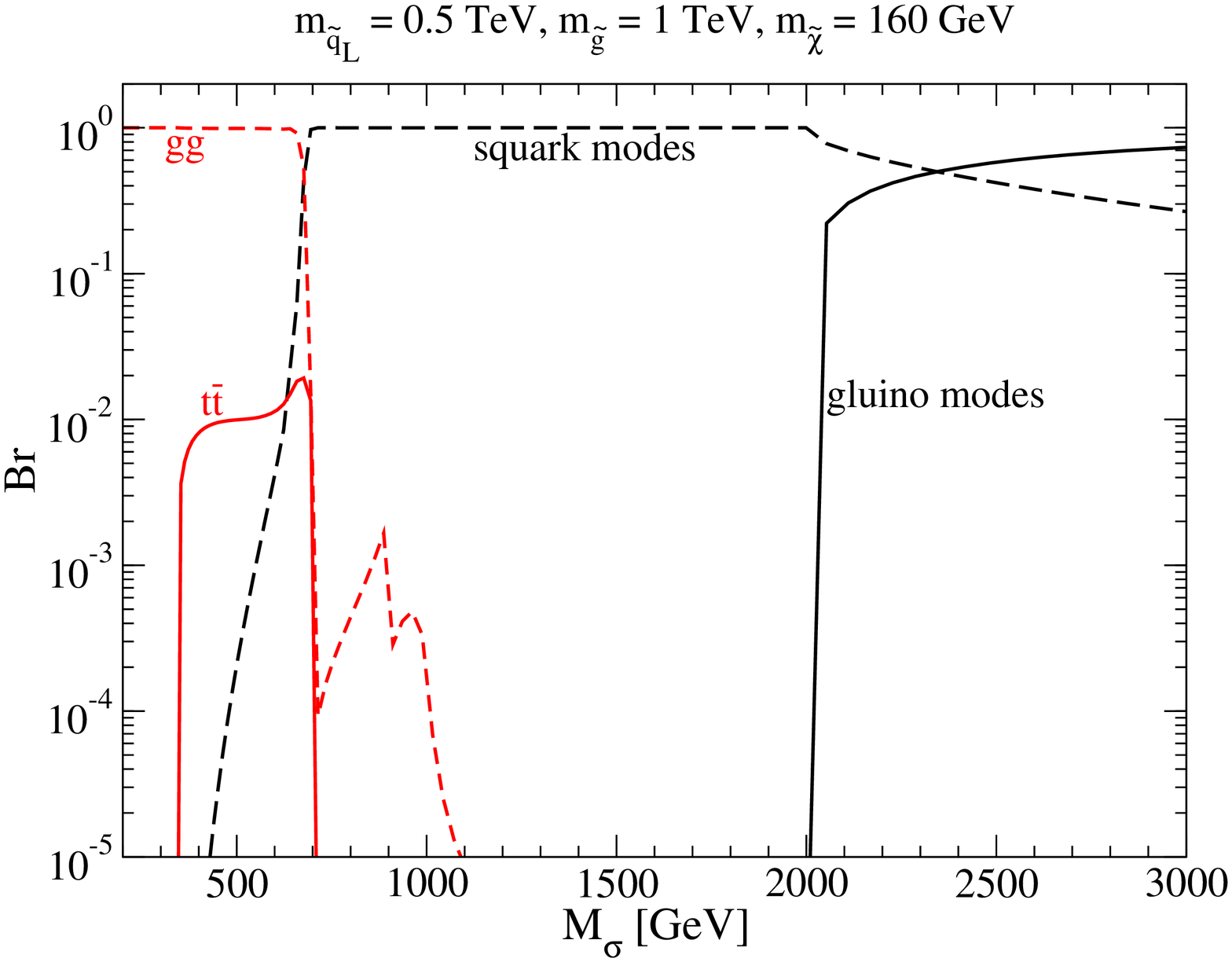, width=7.5cm,height=8.cm}}
\caption{\it Branching ratios for $\sigma$ decays, for $m_{\tilde q_L} = 2
  m_{\tilde g} = 1$ TeV (Left) and $m_{\tilde g} = 2 m_{\tilde q_L} = 1$ TeV
  (Right). In both cases we assumed a neutralino mass $m_{\tilde \chi} =
  0.16 m_{\tilde g}$, and moderate squark mass splitting: $m_{\tilde q_R}
  = 0.95 m_{\tilde q_L}, \, m_{\tilde t_L} = 0.9 m_{\tilde q_L},\,
  m_{\tilde t_R} = 0.8 m_{\tilde q_L}$, with $\tilde t_L$-$\tilde t_R$ mixing
  determined by $X_t = m_{\tilde q_L}$.}
\label{fig:br}
\end{figure}

The corresponding 2-body branching ratios are compared to those for
tree-level decays in Fig.$\,$\ref{fig:br}. Here we assume moderate mass
splitting between the $L$ and $R$ squarks of the five light flavors, and
somewhat greater for soft breaking $\tilde t$ masses: $m_{\tilde q_R} = 0.95
m_{\tilde q_L}, \, m_{\tilde t_L} = 0.9 m_{\tilde q_L},\, m_{\tilde t_R} = 0.8
m_{\tilde q_L}$. We parameterize the off-diagonal element of the squared
$\tilde t$ mass matrix as $X_t m_t$, and take $X_t = m_{\tilde q_L}$. We again
assume the gluino to be a pure Dirac state, i.e. $m_{\tilde g} = |M_3^D|$.

Even for this small mass splitting, the loop decays into two gluons and, if
kinematically allowed, a $t \bar t$ pair always dominate over tree-level
four-body decays $\sigma \rightarrow \tilde g q \bar q \tilde \chi$ (which is
part of the ``gluino modes'' in Fig.~\ref{fig:br}) and $\sigma \rightarrow q
\bar q \tilde \chi \tilde \chi$ (which is part of the ``squark modes''). For
simplicity we evaluated these higher order tree-level decays for a photino
LSP state, with mass $0.16 m_{\tilde g}$. SU(2)$_L$ gauginos have larger
couplings to doublet squarks, but are also expected to be heavier. Including
them in the final state would at best increase the partial widths for
four-body final states by a factor of a few, which would still leave them
well below the widths for the loop induced decays. On the other hand, the
partial width for the tree-level three-body decays $\sigma \rightarrow
\tilde q \bar q \tilde \chi, \ \tilde q^* q \tilde \chi$ can be comparable to
that for the loop-induced decays if $M_\sigma$ is not too much smaller than
$2 m_{\tilde q}$.

Figure~\ref{fig:br} also shows that the ordering between the two loop-induced
decay modes for $M_\sigma > 2 m_t$ depends on the values of various soft
breaking parameters. Increasing the gluino mass increases the $\sigma \tilde q
\tilde q^*$ coupling and hence the partial width into two gluons which is due
to pure squark loops. On the other hand, the $t \bar t$ partial width, which
is due to mixed squark-gluino loops, decreases rapidly with increasing gluino
mass. The increase of the $\sigma \tilde q \tilde q$ couplings is
over-compensated by the gluino mass dependence of the propagators. For
$|M_3^D| > m_{\tilde q}$ the loop functions ${\cal I}_{S,P}$ are additionally
suppressed since then $C_L \simeq C_R, \ D_L \simeq D_R$ up to corrections of
${\cal O}(m^2_{\tilde q} / |M^D_3|^2)$. [A similar cancellation also occurs
for $M_\sigma^2 \gg m^2_{\tilde q}$, for both the $\sigma g g$ and $\sigma t
\bar t$ couplings.] In total, the $t \bar t$ final state will dominate for
small gluino mass and the $gg$ final state for large gluino mass. Moreover, as
noted earlier, the partial width into both gluons and quarks vanishes for
exact degeneracy between $L$ and $R$ squarks.

Not surprisingly, the two-body final states of Eq.$\,$(\ref{Gamma_tree}) that
are accessible at tree level will dominate if they are kinematically allowed.
Note that well above all thresholds the partial width into gluinos always
dominates, since it grows $\propto M_\sigma$ while the partial width into
squarks asymptotically scales like $1/M_\sigma$. This is a result of the fact
that the supersymmetry breaking $\sigma \tilde q \tilde q^*$ coupling has mass
dimension 1, while the supersymmetric $\sigma \tilde g \bar{\tilde g}$
coupling is dimensionless.

\subsection{$\sigma$-Pair Production at the LHC}
\label{subsec:two_sigma}

\begin{figure}[t]
\begin{center}
\begin{picture}(400,120)(0,5)
\Text(0,60)[r]{(a)}
\Text(12,85)[c]{\color{black} $q$}
\Text(12,35)[c]{\color{black} $\bar{q}$}
\ArrowLine(20,85)(50,60)
\ArrowLine(50,60)(20,35)
\Text(65,70)[c]{\color{black} $g$}
\Gluon(50,60)(80,60){2}{6}
\Text(115,85)[l]{\color{black} $\sigma$}
\Text(115,35)[l]{\color{black} $\sigma^\ast$}
\DashArrowLine(80,60)(110,85){3}
\DashArrowLine(110,35)(80,60){3}
\Text(170,60)[r]{(b)}
\Gluon(190,120)(220,95){2}{5}
\Gluon(190,70)(220,95){2}{5}
\Gluon(220,95)(240,95){2}{3}
\Text(255,95)[c]{\color{black} $g_s$}
\DashArrowLine(240,95)(270,120){3}
\DashArrowLine(270,70)(240,95){3}
\Gluon(190,50)(230,50){2}{7}
\Gluon(190,0)(230,0){2}{7}
\DashArrowLine(230,50)(270,50){3}
\DashArrowLine(270,0)(230,0){3}
\DashArrowLine(230,0)(230,50){3}
\Gluon(310,50)(350,50){2}{7}
\Gluon(310,0)(350,0){2}{7}
\DashArrowLine(380,20)(350,50){3}
\DashArrowLine(350,0)(380,30){3}
\DashLine(380,30)(400,50){3}
\DashArrowLine(350,50)(350,0){3}
\DashLine(380,20)(400,0){3}
\Gluon(310,120)(350,95){2}{7}
\Gluon(310,70)(350,95){2}{7}
\Text(350,110)[c]{\color{black} $g^2_s$}
\DashArrowLine(350,95)(390,120){3}
\DashArrowLine(390,70)(350,95){3}
\end{picture}
\end{center}
\caption{\it Feynman diagrams for sigma-pair production in quark annihilation
  (a) and gluon fusion (b).}
\label{fig:feynman}
\end{figure}
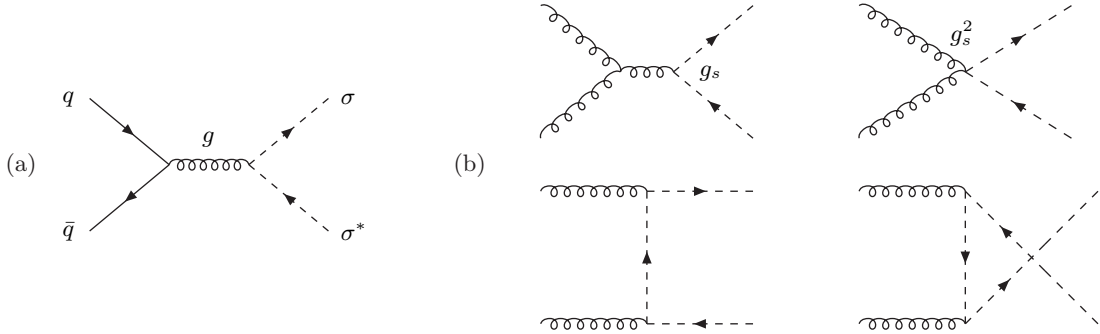

\noindent
As summarized in the preceding section, the phenomenological analysis will be
carried out for a complex color-octet $\sigma$ fields without mass splitting
between the real and imaginary components. The Feynman diagrams for the two
parton processes $gg, q\bar{q} \to \sigma\sigma^{\ast}$ are displayed in
Fig.$\,$\ref{fig:feynman}. They are identical ({\it modulo} color factors) to
squark-pair production \cite{Beenakker,LlSmith} if initial and final-state
flavors are different.

The total cross sections for the two $\sigma\sigma^\ast$ parton processes are
easy to calculate:
\begin{eqnarray}
\sigma [q\bar{q} \to \sigma\sigma^{\ast} ] &=& \frac{4 \pi \alpha_s^2}{9s}
     \,\beta^3_\sigma \,, \\
\sigma [gg \to \sigma\sigma^{\ast} ]
     &=& \frac{15 \pi \alpha_s^2\beta_\sigma}{8s}
     \left[ 1 + \frac{34}{5}\, \frac{M_\sigma^2}{s}-\frac{24}{5}
     \left(1-\frac{M_\sigma^2}{s}\right)\frac{M_\sigma^2}{s}\,
     \frac{1}{\beta_\sigma}
     \log\left(\frac{1+\beta_\sigma}{1-\beta_\sigma}\right)\right]\,.
\end{eqnarray}
The standard notation has been adopted for the parameters: $\sqrt{s}$ is the
invariant parton-parton energy, and $M_{\sigma}$ and $\beta_\sigma =
(1-4M^2_{\sigma}/s)^{1/2}$ the mass and center-of-mass velocity of the
$\sigma$ particle.  The QCD coupling is inserted to leading order,
$\alpha_s(Q^2)= \alpha^{(5)}_s(Q^2) [ 1 + {\alpha^{(5)}_s(Q^2)}/{(6 \pi)}
\cdot \log {M^2_t}/{Q^2} ]^{-1}$, where $\alpha^{(5)}_s(Q^2)$ evolves from
$\alpha^{(5)}_s(M_Z^2) \simeq 0.120$ with $N_F = 5$ flavors by definition,
while the top-quark threshold is accounted for explicitly and supersymmetric
particles do not affect the running in practice; the renormalization scale for
the parton subprocesses is set to $Q = M_{\sigma}$.

While the quark-annihilation cross section increases near threshold with the
third power $\beta^3_\sigma$ of the sgluon velocity, as characteristic for
$P$-wave production, the cross section for equal-helicity gluon-fusion
increases steeply $\sim \beta_\sigma$ with the velocity, as predicted for
$S$-waves by the available phase space. Asymptotically
the two parton cross sections scale $\propto s^{-1}$.

\begin{figure}[t]
\epsfig{figure=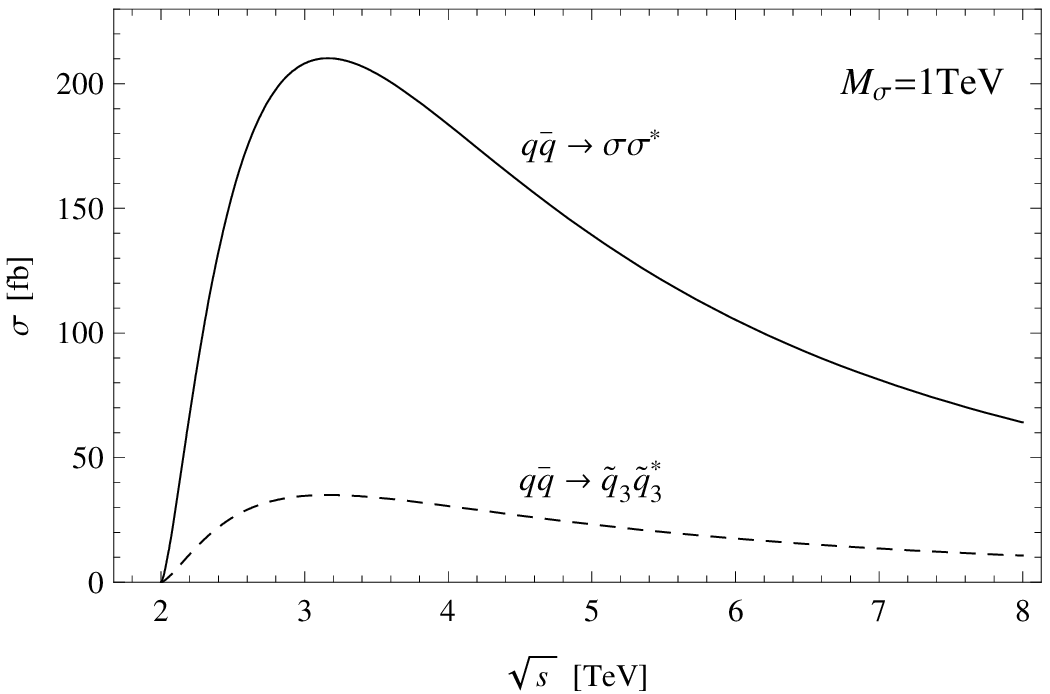, height=6.5cm, width=7cm}
\hspace{1cm}
\epsfig{figure=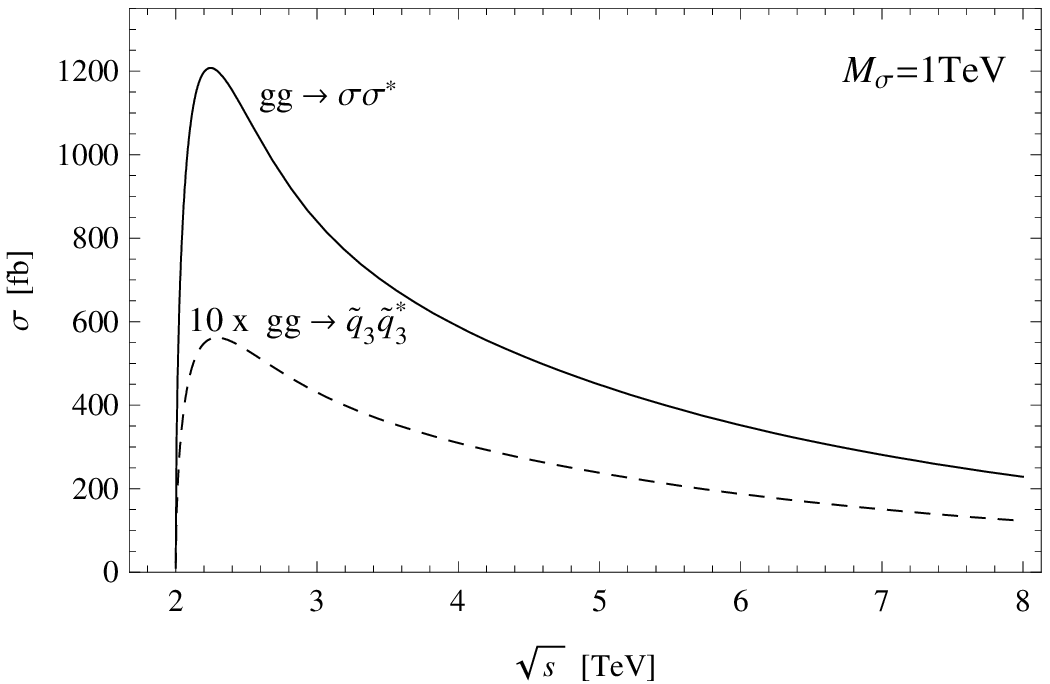, height=6.5cm, width=7cm}
\caption{\it Parton cross sections for $\sigma\sigma^\ast$ production in the
         $q\bar{q}$ (Left) and $gg$ (Right) channel. For comparison, the
         production of 3rd generation squark pairs is shown by the dashed
         lines for the same masses.  }
\label{fig:crosssections}
\end{figure}

The $\sigma{\sigma}^\ast$ cross sections are compared in
Fig.$\,$\ref{fig:crosssections} with the production of squark pairs [of the
3rd generation to match the dynamical production mechanisms]: $gg, q\bar{q}
\to \qw_3 \qw_3^{\ast}$. As expected, the $\sigma\sigma^\ast$ cross sections
exceed the $\qw_3 \qw_3^{\ast}$ cross sections by a large factor, i.e. $\sim
20$ for $gg$ collisions and 6 for $q\bar{q}$ collisions. This can be
exemplified by considering the evolution of ratios for the cross sections
from small to maximum velocity, $\beta$ being again
the center-of-mass velocity of the sgluon or squark in the
final state:
\begin{eqnarray}
\frac   {\sigma \left[ gg \to \sigma \sigma^{\ast} \right] }
{   \sigma \left[ gg \to \qw_3 \qw^{\ast}_3 \right]} & = &
\left\{ \begin{array}{cl}
 \displaystyle{
      \frac { \tr \left[ \{F^a, F^b\} \{F^a, F^b\} \right]}
      { \tr \left[ \left\{\frac{\lambda_a}{2}, \frac{\lambda_b}{2} \right\}
\left\{\frac{\lambda_a}{2}, \frac{\lambda_b}{2} \right\} \right] } }=
\frac{216}{28/3} \simeq 23  & \ \ {\rm for}\ \beta\to 0\,, \\
 & \\
\displaystyle{
\frac{ \tr\,(2 F^a F^b F^b F^a + F^a F^b F^a F^b ) }
       {  \tr\left(2 \frac{\lambda^a}{2} \frac{\lambda^b}{2}
                            \frac{\lambda^b}{2} \frac{\lambda^a}{2}
+ \frac{\lambda^a}{2} \frac{\lambda^b}{2} \frac{\lambda^a}{2}
       \frac{\lambda^b}{2} \right) } } = \frac {180}{10} = 18
& \ \ {\rm for}\ \beta \to 1\,,
\end{array} \right.
\label{eq:ratio_af}
 \\[2mm]
\frac{   \sigma \left[ q\bar{q} \to \sigma \sigma^{\ast} \right] }
{   \sigma \left[ q\bar{q} \to \qw_3 \qw^{\ast}_3 \right]}
      &=& \frac{\tr\left(\frac{\lambda^a}{2}\,\frac{\lambda^b}{2}\right)\,
                \tr\left(F^a F^b\right)}
                {\tr\left(\frac{\lambda^a}{2}\,\frac{\lambda^b}{2}\right)\,
                \tr\left(\frac{\lambda^a}{2}\,\frac{\lambda^b}{2}\right)}
                 = \frac{12}{2} = 6 \quad \mbox{for any}\ \ \beta\,.
\end{eqnarray}
The ratio (\ref{eq:ratio_af}) decreases monotonically as $\beta$
increases but by no more than 20\%. Most important is the ratio at the
maximum of the $gg$ cross sections where it is still close to the initial
maximal value; this can easily be explained by observing that, in Feynman
gauge, the leading contribution is generated by the quartic coupling.
The differences in the color factors reflect the different strengths
of the couplings in the fractional triplet $\lambda /2$ and the integer octet
$F$ couplings of SU(3)$_C$ with $(F^a)_{bc}=-i f^{abc}$. The cross
sections are shown in Fig.$\,$\ref{fig:crosssections} for $M_{\sigma}=1$ TeV
across the invariant energy range relevant for the LHC. The values of the
maxima in the $gg$ and $q\bar{q}$ channels are about 1 pb and 0.2 pb,
respectively, a typical size for such processes.

\begin{figure}[h]
\vskip 0.5cm
\epsfig{figure=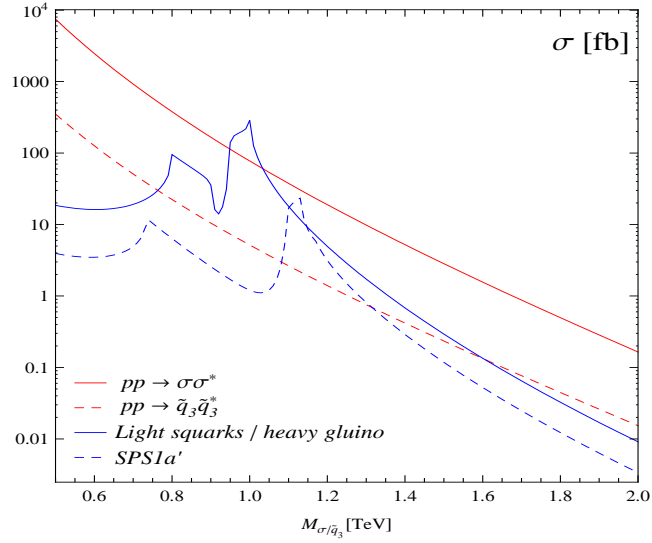, height=7.2cm, width=8.5cm}
\caption{\it Cross sections for $\sigma$-pair [and $\tilde{q}_3$-pair]
         production (red lines), as well as for single $\sigma$ production (blue lines),
         at the LHC. In the latter case the solid blue curve has been obtained using
         the same mass parameters as in Fig.2 (Right), while the dashed blue
         curve adopts the mSUGRA benchmark point SPS1a$'$.}
\label{fig:lhcsigma}
\end{figure}

The cross section for $\sigma$-pair production at LHC, $pp \to \sigma
\sigma^\ast$, is shown by the solid red curve in Fig.$\,$\ref{fig:lhcsigma}
for the $\sigma$-mass range between 500 GeV and 2 TeV [adopting the LO CTEQ6L
parton densities \cite{CTEQ}]. The cross section exceeds stop or sbottom-pair
production (red dashed line), mediated by a set of topologically equivalent
Feynman diagrams, by more than an order of magnitude, as anticipated at the
parton level. With values from several picobarn downwards, a sizable
$\sigma\sigma^\ast$ event rate can be generated.

With the exception of $\sigma \rightarrow g g$ decays, all the modes shown in
Fig.$\,$\ref{fig:br} give rise to signatures that should be easily detectable
if $\sigma$ is not too heavy. The most spectacular signature results from
$\sigma \rightarrow \tilde g \tilde g$ decay, each $\sigma$ decaying into at
least four hard jets and two invisible neutralinos as LSP's. $\sigma$-pair
production then generates final states with a minimum of eight jets and four
LSP's, as noted in the Introduction.

The transverse momenta of the hard jets produced in the simplest case $\tilde
\chi = \tilde \chi_1^0$ can easily be estimated by analyzing production and
decays near the mass thresholds, i.e. $M_{\sigma} \simeq 2 m_{\gw} \simeq 2
m_{\qw} \gg m_{{\tilde{\chi}}^0_1}$. In this kinematic configuration the total
jet transverse energy and the average jet transverse energy amount to
\begin{equation}
\sigma\sigma^\ast
    \; : \;\;\;
\langle {E_{{\perp}j}^{tot}} \rangle \sim 2 m_{\qw}  \;\;\; {\rm{and}} \;\;\;
\langle {E_{{\perp}j}} \rangle       \sim m_{\qw} / 4    \;.
\end{equation}
The total transverse energy $E_T$ carried by the LSPs and the vector sum of the
momenta of the four $\tilde{\chi}^0_1$ in the final state, which determines
the measured missing transverse momentum $p_T$, are predicted to be
\begin{equation}
\sigma\sigma^\ast
    \; : \;\;\;
\langle {E_{{\perp}{\tilde\chi}}^{tot}} \rangle \sim 2 m_{\qw} 
\;\;\; {\rm{and}}
    \;\;\; \langle {p_{{\perp}{\tilde\chi}}} \rangle \sim m_{\qw}
\end{equation}
in the random-walk approximation for the $\tilde\chi$ momenta in the
transverse plane. This is to be contrasted to gluino-pair production near
threshold, where the corresponding observables are for the same mass
configuration:
\begin{eqnarray}
\gw\gw
  \; &:& \;\;\;
\langle {E_{{\perp}j}^{tot}} \rangle \sim m_{\qw}  \;\;\; {\rm{and}} \;\;\;
\langle {E_{{\perp}j}} \rangle \sim m_{\qw} / 4\,,    \\
&& \;\;\; \langle {E_{{\perp}{\tilde\chi}}^{tot}} \rangle \sim m_{\qw}
          \;\;\; {\rm{and}} \;\;\;
\langle {p_{{\perp}{\tilde\chi}}} \rangle \sim m_{\qw} / \sqrt{2}  \;.
\end{eqnarray}
Thus, the total jet transverse energies and the missing transverse momenta are
markedly different in the N=1 and N=2 theories for the same mass
configurations.

These simple estimates are backed up by a Monte-Carlo simulation of $\sigma$-pair
production at the LHC, followed by the decay into four on-shell gluinos.
The total transverse jet energy and the vector sum of the LSP transverse
momenta are summarized in Tab.$\,$\ref{tbl:transspec} for a spectrum of
$\sigma$-masses, and fixed ratios of gluino, squark and LSP neutralino masses.
The squark and gluino masses are again chosen at about half a TeV. The values
of the transverse momenta match the earlier estimates quite well. It should be
noted however that the jet transverse momenta fall into two groups. The
transverse momenta of jets in gluino to squark decays are generally small
while the transverse momenta of the jets generated in squark decays are large.
Both groups are populated equally so that the average transverse momenta of
the jets are reduced by an approximate factor two compared with the MSSM
gluino pair production [setting $m_{\tilde g}|_{\rm MSSM} = M_\sigma|_{\rm hybrid
  \ model}$ for the proper comparison].

\begin{table}[ht!]
\centering
\caption{\it Transverse jet energies and vector sum of the LSP transverse momenta
         for final states in $2\sigma$ and $2\gw$ production, with primary
         $\sigma / \gw$-masses of 1.5 and 0.75 TeV; the mass hierarchy in
         the cascade decays is noted in the bottom line. Below the
         transverse energy per jet of the total jet ensemble [tot], the
         transverse energies in the high and the low jet-energy groups
         [high/low] are displayed. All quantities in TeV.}
\label{tbl:transspec}
\vskip 0.5cm
\begin{tabular}{|rl||cc|cc|cc|}
\hline
\multicolumn{ 2}{|c||}{$M_{\sigma / \gw}$} & \multicolumn{ 2}{|c|}{2$\sigma$}
      & \multicolumn{ 2}{|c|}{2$\gw$} &    2$\sigma$ &        2$\gw$ \\
\cline{3-8}
\multicolumn{ 2}{|c||}{} & $\langle E_{\perp j}^{tot}\rangle $
      & $\langle E_{\perp j}\rangle$ & $\langle E_{\perp j}^{tot}\rangle$
      & $\langle E_{\perp j}\rangle$ & {$\langle p_{\perp \tilde\chi}\rangle$}
      & {$\langle p_{\perp \tilde\chi}\rangle$}\\
\hline \hline
  1.50 TeV &  [tot]  & 1.67 & 0.21 & 1.67 & 0.42 & 0.45 & 0.65 \\

           &  [high] &      & 0.27 &      & 0.53 &      &      \\

           &  [low]  &      & 0.15 &      & 0.31 &      &      \\
\hline
  0.75 TeV &  [tot]  & 0.91 & 0.11 & 0.93 & 0.23 & 0.22 & 0.31 \\

           &  [high] &      & 0.14 &      & 0.29 &      &      \\

           &  [low]  &      & 0.08 &      & 0.17 &      &      \\
\hline
\multicolumn{ 8}{|l|}{ $M_{\sigma} = 2\, M_{\gw} = 8/3 \,
           M_{\qw} = 15\, M_{\tilde\chi}$} \\
\hline
\end{tabular}
\end{table}
\vskip 0.2cm

Other interesting final states resulting from $\sigma$-pair production are
four-stop states $\tilde t_1 \tilde t_1 \tilde t_1^* \tilde t_1^*$, which can
be the dominant mode if $m_{\tilde q} \lesssim m_{\tilde g}$ and $L$-$R$
mixing is significant in the stop sector, and $\tilde q \tilde q^* \tilde g
\tilde g$, which can be a prominent mode if $M_{\sigma} > 2 m_{\tilde g}
\gtrsim 2 m_{\tilde q}$.  These channels also lead to four LSPs in the final
state, plus a large number of hard jets. On the other hand, the $t t \bar t
\bar t$ final state, which can be the dominant mode if the two-body decays
into squarks and gluinos are kinematically excluded, might allow the direct
kinematic reconstruction of $M_\sigma$.

\subsection{Single $\sigma$ Channel}
\label{subsec:single_sigma}

\noindent
As noted earlier, sgluons can be generated singly in gluon-gluon collisions
via squark loops. The partonic cross section, with the Breit-Wigner function
factorized off, is given by
\begin{equation} \label{sig_gg}
{\hat{\sigma}} [ gg \to \sigma ] = \frac{\pi^2} {M^3_\sigma} \Gamma(\sigma \to
gg)\,,
\end{equation}
where the partial width for $\sigma \to gg$ decays has been given in
Eq.$\,$(\ref{Gamma_gg}).

The resulting cross section for single $\sigma$ production at the LHC is shown
by the blue curves in Fig.$\,$\ref{fig:lhcsigma} [based on the LO CTEQ6L parton
densities \cite{CTEQ}]. The solid curve has been calculated for the parameter
set of the right frame of Fig.$\,$\ref{fig:br}, while the dashed curve has
been determined by taking the soft breaking parameters in the gluino and
squark sector from the widely used benchmark point SPS1a$'$ \cite{sps}. In the
former case the single $\sigma$ cross section can exceed the $\sigma$-pair
production cross section for $M_\sigma \sim 1$ TeV. Since SPS1a$'$ has a
somewhat smaller gluino mass [which we again interpret as a Dirac mass here]
it generally leads to smaller cross sections for single $\sigma$ production.
Taking $m_{\tilde q} \simeq 2 |M_3^D|$, as in the left frame of Fig.~2, would
lead to a very small single $\sigma$ production cross section. Recall that
$m_{\tilde q} > |M_3^D|$ is required if $\sigma \to t \bar t$ decays are to
dominate. We thus conclude that one cannot simultaneously have a large
$\sigma(pp \to \sigma)$ and a large ${\rm Br}(\sigma \to t \bar t)$.

The signatures for single $\sigma$ production, which is an ${\cal
  O}(\alpha_s^3)$ process, are potentially exciting as well. However, since
all final states resulting from $\sigma$ decay can also be produced directly
in tree-level ${\cal O}(\alpha_s^2)$ processes at the LHC, it is a problem to
be solved by experimental simulations whether single $\sigma$ production is
detectable as a resonance above the SM plus MSSM backgrounds, given that in
most cases, with the exception of the 2-gluon channel, the direct kinematic
reconstruction of $M_\sigma$ is not possible.

\section{SUMMARY}
\label{sec:summary}

\noindent
The color-octet scalar sector in the N=1/N=2 hybrid model we have analyzed in this letter,
leads to spectacular signatures of supersymmetry which are distinctly
different from the usual MSSM topologies. Depending on the masses of the
particles involved, either multi-jet final states with high sphericity and
large missing transverse momentum are predicted, or four top quarks should be
observed in $2\sigma$ production.  If the mass splitting between $L$ and $R$
squarks is not too small, loop-induced single $\sigma$ production may also
have a sizable cross section; however, this channel suffers from much larger
backgrounds, though identifying the $\sigma$ particle as a resonance in
2-gluon final states would truly be an exciting experimental observation.

In this letter we assumed that gluinos are pure Dirac states, and that the two
components of the complex scalar field, $\sigma = (S+iP)/\sqrt{2}$, are
degenerate. Relaxing these assumptions would introduce more parameters into
the scheme, yet the central characteristics of the experimental event
topologies of the final states at LHC would not change significantly. For
example, for fixed mass, the $S$ or $P$ pair production cross section is
simply half the $\sigma$ pair production cross section.

%
\acknowledgments{}

\noindent
The work by SYC was supported by the Korea Research Foundation Grant funded
by the Korean Government (MOERHRD, Basic Research Promotion Fund)
(KRF-2008-521-C00069). The work of MD was partially supported by
Bundesministerium f\"ur Bildung und Forschung under contract no. 05HT6PDA,
and partially by the Marie Curie Training Research Networks ``UniverseNet''
under contract no. MRTN-CT-2006-035863, ``ForcesUniverse'' under contract no.
MRTN-CT-2004-005104, as well as ``The Quest for Unification'' under contract
no. MRTN-CT-2004-503369. JK was supported by the Polish Ministry of Science
and Higher Education Grant no~1~P03B~108~30 and the EC Programme``Particle
Physics and Cosmology: the Interface'' under contract no.
MTKD-CT-2005-029466. PMZ is grateful for the warm hospitality extended to
him at the Inst.~Theor.~Phys.~E of RWTH Aachen, and at LPT/Orsay of Paris-Sud.
We are particularly thankful to T.~Plehn and T.~Tait for their cooperation in
clarifying the origin of discrepancies between their first version
Ref.~\cite{TT} with our results.

\vskip 0.3cm
%

\end{document}